\documentclass[preprint,floats,aps,epsfig,nofootinbib,amssymb]{revtex4-1}
\usepackage{mathrsfs}

\usepackage{slashed}
\usepackage{graphicx,color}
\usepackage{dcolumn}
\usepackage{bm}
\usepackage{subfig}
\usepackage{graphicx}
\usepackage{amssymb}

\usepackage[colorlinks]{hyperref}
\hypersetup{
pdfauthor={Y. Cai, W. Chao},
pdftitle={The Higgs Seesaw Induced Neutrino Mass and Dark Matter}
}

\begin{document}

\preprint{ACFI-T14-15}

\title{The Higgs Seesaw Induced Neutrino Masses and Dark Matter }

\author{Yi Cai$^1$}

\email{yi.cai@unimelb.edu.au}

\author{Wei Chao${^{2,3}}$}
\email{chao@physics.umass.edu}

\affiliation{$^1$ARC Centre of Excellence for Particle Physics at the Terascale,
School of Physics, The University of Melbourne, Victoria 3010, Australia\\
  $^2$Amherst Center for Fundamental Interactions, Department of Physics, University of Massachusetts-Amherst
Amherst, MA 01003 \\
$^3$INPAC, Shanghai Jiao Tong University, Shanghai, China
 }

\begin{abstract}
In this paper we propose a possible explanation of the active neutrino Majorana masses with the TeV scale new physics which also provide a dark matter candidate. We extend the Standard Model (SM) with  a local $U(1)'$ symmetry and introduce a seesaw relation for the vacuum expectation values (VEVs) of the exotic scalar singlets, which break the $U(1)^\prime$ spontaneously.  
The larger VEV is responsible for generating the Dirac mass term of the heavy neutrinos, while the smaller for the Majorana mass term.
%The larger and smaller VEVs are responsible for generating the Dirac and Majorana mass terms of the heavy neutrinos respectively. 
As a result active neutrino masses are generated via the modified inverse seesaw mechanism. The lightest of the new fermion singlets,
which are introduced to cancel the $U(1)^\prime$ anomalies, can be a stable particle with ultra flavor symmetry and thus a plausible dark matter candidate. 
We explore the parameter space with constraints from the dark matter relic abundance and  dark matter direct detection.

%In this work we propose a natural explanation of the small Majorana mass terms of heavy neutrinos in the inverse seesaw model.   
%We extend the inverse seesaw model with three pairs of fermion singlets and two Higgs singlets as well as a local $U(1)'$ symmetry.  
%After the spontaneously breaking of the electroweak gauge symmetry and the extra $U(1)'$, 
%the two Higgs singlets get vacuum expectation values (VEVs), 
%of which the larger and the smaller one are responsible for the Dirac and Majorana mass terms of the the heavy neutrinos respectively.
 %As a result, the inverse seesaw mechanism is naturally realized.  
%Meanwhile, the new gauge singlet introduced to cancel the anomalies of the $U(1)'$ symmetry is a plausible dark matter candidate.  
%We study its implication in dark matter relic abundance and direct detections.
\end{abstract}

\draft

\maketitle
\section{introduction}

With the discovery of the Higgs-like scalar  at the CERN LHC, the Standard Model Higgs mechanism for spontaneous breaking of the $SU(2)_{\rm L}\times U(1)_{\rm Y}$ gauge symmetry appears to be a correct description of the nature. In addition to explaining the spontaneous breaking of the electroweak symmetry, the Higgs boson is also responsible for the origin of fermion masses, via the Yukawa interactions.  
On the other hand, the minimal Higgs mechanism is not able to address the 
%On the other hand, the minimal Higgs mechanism is more or less incomplete,  considering 
the fermion mass hierarchy problem, where the quark-lepton masses range from the top quark with mass of order electroweak scale, $M_t=172~{\rm GeV}$, down to electron of mass, $M_e=0.511$ MeV, and  the first order phase transition, relevant for baryon asymmetry of the Universe.  More precise measurement of Higgs boson properties will help determine whether there are new degrees of freedom that participate in electroweak symmetry breaking
or otherwise involve in  new Higgs boson interactions.

Furthermore, the discovery of the neutrino oscillation has confirmed the theoretical
expectation that neutrinos are massive and lepton flavors are mixed~\cite{pdg},
which provided the first piece of evidence for physics beyond the Standard Model
(SM).  In order to accommodate the tiny neutrino
masses, one can extend the SM by introducing several right-handed
neutrinos, which are taken to be singlets under the $SU(2)_{\rm
L}^{} \times U(1)_{\rm Y}^{}$ gauge group. In this case, the gauge
invariance allows right-handed neutrinos to have Majorana mass
$M_R^{}$, which is not subject to the electroweak symmetry breaking
scale. Thus the effective mass matrix of three light Majorana
neutrinos can be highly suppressed if $M_R^{}$ is much larger than the electroweak scale, 
which is the so-called canonical seesaw mechanism~\cite{seesawI}. 
Two other types of tree-level seesaw mechanisms have also been proposed~\cite{seesawII, seesawIII}.
Despite its simplicity and elegance, the canonical seesaw mechanism is impossible to be tested 
in current collider experiments, especially at the Large Hadron Collider, 
due to its inaccessibly high right-handed Majorana mass scale.
% suffers from two serious problem. 
%First of all, it is impossible to test these models in current collider experiments.
%First, the testability of such seesaw models is impossible in collider
%experiments, especially in the CERN Large Hadron
%Collider (LHC). 
Heavy Majorana neutrinos can also give large radiative corrections to the SM Higgs mass,
which causes the seesaw hierarchy problem~\cite{biggio}. 
%In TeV scale seesaw models~\cite{Tevseesaw,tevseesawrev},
%the new physics is manifest at the TeV scale, which in principle can be tested 
%at the LHC and the seesaw hierarchy problem is eased. 
%Light neutrino masses, however, receive radiative corrections from different 
%heavy Majorana neutrinos. These contributions have to be fine-tuned to cancel each other,
%which makes TeV scale seesaw models unnatural.    
%Second, the SM Higgs mass is very sensitive to
%quantum corrections from the heavy Majorana neutrinos, which induces
%the seesaw hierarchy problem\cite{biggio}. To submit to the experiment, several
%In TeV scale seesaw models\cite{Tevseesaw,tevseesawrev} were proposed, where fine-tunings of
%cancellations among the contributions to light neutrino masses from
%different heavy Majorana neutrinos have to be employed. Such models
%suffer from dangerous radiative corrections and looks unnatural.
An alternative way to generate tiny Majorana neutrino masses at the TeV scale 
is the inverse seesaw mechanism~\cite{inverseseesaw,invseesaw2}, 
in which the neutrino mass $m_\nu$ is proportional to a small effecitive Majorana mass term $\mu$.
%It's characterized by a small effective Majorana mass term $\mu$ such that $m_\nu\sim \mu$. 
But there is no dynamical explanation of the smallness of $\mu$.
The argument is that neutrinos become massless in the limit of vanishing $\mu$ 
and the global  lepton number, $U(1)_{\rm L}$,  is then restored, leading to a larger symmetry \cite{thooft}. 
This argument, however, only works when we give the left-handed singlets ($S_L^{}$) the same quantum(lepton) number as that of  the right-handed heavy neutrinos ($N_R^{}$).  
If the lepton number of $S_L^{} $ is zero, the argument above does not hold up any more.  
Besides, the lepton number is only an accidental symmetry of the SM and is explicitly 
broken by anomalies.

Since neutrino is the only neutral matter field in the SM, it is reasonable to conjecture that neutrinos are correlated with the dark matter, which provides another evidence of the new physics beyond the SM from the precise cosmological observations, through certain dark symmetry. The nature of the dark matter and the way it interacts with ordinary matter are still mysteries. The discovery of the Higgs boson opens up new ways of probing the world of the dark matter.  The neutrino flux from the annihilation of the dark matter at the center of the dark matter halo also provides a way of indirect detecting the dark matter.

In this paper, we propose a possible explanation of the smallness of neutrino masses and a possible candidate of the dark matter.  
%{\color{red} 
The discovery of the Higgs-like boson makes the Higgs mechanism more promising as a possible way to understand the origin of
the fermion masses. We study the possibility of generating a small Majorana mass  term with
the help of the  seesaw mechanism in the Higgs sector. 
%As a result, the Higgs mechanism becomes more and more reality.  Perhaps the Higgs mechanism is the only correct way to understand the origin of all fermion masses. Here we study the possibility of explaining the smallness of $\mu$ with the help of Higgs mechanism.  
%I am not sure if this argument is convincing.
%}
We extend the SM with a local $U(1)'$ gauge symmetry,
which is spontaneously broken by the vacuum expectation value (VEV) $\langle \varphi \rangle$ of an extra scalar singlet.  Furthermore there is a seesaw mechanism in the scalar singlet sector: a second scalar singlet gets a tiny VEV $\langle \Phi \rangle$ in a way similar to that of the Higgs triplet in the type-II seesaw model~\cite{seesawII}.  
$\langle \varphi \rangle $ is responsible for the origin of the dark matter mass and the Dirac neutrino mass term, 
while $\langle \Phi \rangle $ is responsible for the origin of a small Majorana neutrino mass term.   The active neutrino mass matrix arises  from the modified inverse seesaw mechanism.  A crucial feature of our model is that all the mass terms originate from the spontaneous breaking of  local gauge symmetries, and dark matter is correlated with the neutrino physics via the $U(1)'$ gauge symmetry. We study constraints on the parameter space of this model from astrophysical observation and dark matter direct detections.

The paper is organized as follows: In section II we describe our model, including the full Lagrangian, Higgs VEVs and mass spectrum. In section III we study the neutrino masses and the effective lepton mixing matrix of the model. 
Section IV is devoted to the study of the dark matter phenomenology.  We summarize in section V. 

\section{The model}
We extend the SM with three generations of right-handed neutrinos $N_R$ and singlets $S_L$ as in  the inverse seesaw mechanism, together with two extra scalar singlets, $\varphi$ and $\Phi$, as well as a spontaneously broken $U(1)'$ gauge symmetry and a global $U(1)_D$ flavor symmetry. 
The quantum numbers of the fields are given in Table~\ref{tab:qn}, 
where $\ell_L$ is left-handed lepton doublet, 
$E_R^{}$ is the right-handed charged lepton, 
$H$ is the SM Higgs doublet, 
and $\chi_L^{}$ and $\chi_R^{}$ are the fermion singlet pair carrying the same $U(1)_D$ quantum number.
Three generations of gauge singlets $\chi_{L,R}$  are needed 
to cancel anomalies~\cite{globalsu2,avector1,avector2,avector3,anog1,anog2,anog3} of the $U(1)'$ gauge symmetry.   
The lightest generation  of $\chi_{L,R}$ is stable due to the global $U(1)_D$ flavor symmetry and thus plays the role of 
dark matter~\cite{dmn1,dmn2,dmn3}.

\begin{table}[htbp]
\centering
\begin{tabular}{c|c|c|c|c|c|c ||c|c|c}
\hline \hline ~~~~~~~~~~ 
& $\ell_L^{}$ &$E_R^{}$& $ N_R^{}$ & $S_L^{}$&$\chi_R^{}$&$\chi_L$ & H & $\varphi $ & $\Phi$\\
\hline$U(1)'$&$0$ & $0$ &$0$ &$1$ & $1$ & 0  &$0$& $1$ & $2$ \\
\hline
$U(1)_D$ &0&0&0&0&$1$&$1$&0 &0&0\\
\hline
\hline
\end{tabular}
\caption{ Quantum numbers of the relevant fields under the local $U(1)'$ and the global $U(1)_D$ flavor symmetry.   }\label{tab:qn}
\end{table}

The Higgs potential of the model can be written as
\begin{eqnarray}
\nonumber
V= &&-m^2 H^\dagger H  - m_1^2 \varphi^\dagger \varphi + m_2^2 \Phi^\dagger \Phi + \lambda (H^\dagger H)^2 + \lambda_1^{} (\varphi^\dagger \varphi)^2  + \lambda_2^{} (\Phi^\dagger \Phi)^2 \\
\nonumber
& & + \lambda_3^{} (H^\dagger H) (\varphi^\dagger \varphi )   + \lambda_4^{} (H^\dagger H ) (\Phi^\dagger \Phi) + \lambda_5^{} (\varphi^\dagger \varphi ) (\Phi^\dagger \Phi ) \\
&& +\sqrt{2}\lambda_6^{}   \left( \Lambda \varphi^2 \Phi + {\rm h.c. }\right) \; , \label{potential}
\end{eqnarray}
where we define $H= (h^+, ~(h_0+i A + v) /\sqrt{2})^T$, $\varphi =(\varphi_0+ i \delta + v_1^{} ) /\sqrt{2}$ and $\Phi = (\Phi_0^{} + i \rho + v_2^{}) /\sqrt{2}$. After imposing the conditions of the global minimum, one has
\begin{eqnarray}
&&- m^2 v+ \lambda v^3 + {1 \over 2 } v(\lambda_3^{} v_1^2 + \lambda_4^{} v_2^2 ) =0 \;  ,\\
&& -m_1^2 v_1 + \lambda_1^{} v_1^3 + {1 \over 2 } v_1^{} ( \lambda_3^{} v^2 + \lambda_5 v_2^2 ) + 2\lambda_6^{} \Lambda v_2^{}  =0 \; , \\
&&+ m_2^2 v_2 + \lambda_2^{} v_2^3 + {1 \over 2} v_2^{} (\lambda_4^{} v^2 + \lambda_5^{} v_1^{2} ) +  \lambda_6^{} \Lambda v_1^2  =0 \; .
\end{eqnarray}
Then the VEVs can be solved in terms of the parameters
\begin{eqnarray}
v^2\approx { 2 m_1^2 \lambda_3^{} - 4 m^2 \lambda_1 \over \lambda_3^2 -4 \lambda_1^{} \lambda }  \; ,\hspace{1cm}
v_1^2 \approx { 2m^2 \lambda_3 - 4  m_1^2 \lambda \over \lambda_3^2 - 4 \lambda \lambda_1^{} } \; , \hspace{1cm} v_2^{} \approx -{2 \lambda_6 \Lambda v_1^2 \over 2 m_2^2 + \lambda_4^{} v^2 + \lambda_5^{} v_1^2 } \; , \label{vev}
\end{eqnarray}
where $v_2^{}$ is proportional to $\Lambda$  and suppressed by  $m_2^{2}$. Thus $v_2{}$ can be a small value given a  large $m_2^{2}$ or small $\Lambda$.

In the basis $(h_0, ~\phi_0, ~\Phi_0)$, the mass matrix of the CP-even Higgs can be written as
\begin{eqnarray}
M_{\rm CP even}^2 = \left( \matrix{ 2 v^2 \lambda & vv_1 \lambda_3 & vv_2 \lambda_4 \cr vv_1^{} \lambda_3 & 2\lambda_1^{} v_1^2 & 2\Lambda v_1^{} \lambda_6 \cr vv_2 \lambda_4 & 2 \Lambda v_1 \lambda_6 &2 v_2^2 \lambda_2 -\lambda_6 \Lambda v_1^2 v_2^{-1} } \right) \; .
\label{cpeven}
\end{eqnarray}
The mass eigenstates of the CP-even Higgs are then denoted as $h_i$ including the SM-like Higgs $h$ and two exotic Higgs, $h_1$ and $h_2$.
There is no mixing between the SM CP-odd Higgs $A$, which is the Goldstone boson eaten by the $Z$ gauge boson, and those of the Higgs singlets,  i.e. $\delta$ and $\rho$.  The mass matrix of the CP-odd Higgs singlets in the basis of 
$(\delta, \rho)$ is 
\begin{eqnarray}
M^2_{\rm CP-odd} = \left( \matrix{-4 \Lambda v_2^{} \lambda_6 & 2 \Lambda v_1^{} \lambda_6 \cr 2 \Lambda v_1 \lambda_6 & -\Lambda \lambda_6 v_1^2 v_2^{-1}}\right)\; . \label{odd}
\end{eqnarray}
The massless eigenstate  of  the eq. (\ref{odd}) is the Goldstone boson eaten by the $Z^\prime$
%Here $\delta$ is the Goldstone boson eaten by the $Z^\prime$. 
and the nonzero mass eigenstate of the CP-odd scalar 
is then denoted as $A'$, the mass squared of which can be written as  
$m_{A^\prime}^2 = -4 (v_2 + v_1^2 v_2^{-1} ) \Lambda \lambda_6$.

Since the SM particles are not charged under $U(1)'$, there is no experimental constraint on the new symmetry. 
Besides, there is no tree-level mixing between $Z$ and $Z^\prime$. Thus the mass and coupling constant of $Z^\prime$ 
are not constrained by current experiments either. 
%Since the SM fields carry no $U(1)$ charge, we n to wary about  the experimental constraint on the new symmetry. Besides there is no tree level mixing between the $Z$ and $Z^\prime$. As a result,  the mass and coupling constant of $Z^\prime$ are quite arbitrary.  

\section{Neutrino Masses}
Now we investigate how to realize the neutrino masses  in our model. 
The Yukawa interactions of the lepton sector can be given by
\begin{eqnarray}
-\mathcal{L}= \overline{\ell_L^{}} Y_E^{} H E_R^{} + \overline{\ell_L^{} } Y_\nu^{} \tilde{H} N_R^{} + \overline{S_L^{} } Y_N^{} \varphi N_R^{} + \overline{S_L^{} } Y_S^{} \Phi S_L^C + \overline{\chi_L^{}} Y_\chi^{} \varphi \chi_R^{} + {\rm h.c.}
\end{eqnarray}
where the first and second terms are the charged lepton  and neutrino Yukawa interactions separately, the third and fourth terms are the Yukawa coupling of heavy neutrinos to the scalar singlets, 
and the last term is the Yukawa coupling of the additional fermions.  
We assume that there is no  $ \overline{N_R^C} M N_R^{}$ type of mass term,
which can be easily forbidden by an extra global $U(1)$ symmetry, 
in which all the right-handed fermions, $H$ and $S_L$ are singly charged, 
$\Phi$ doubly charged and all other particles neutral.
%under which the hyper-charge  of right-handed fermions, $H$ and $S_L^{}$ is $1$,  the hyper-charge  $\Phi$ is $2$, while all the other fields charge zero.  
The symmetry is explicitly broken by the last term of the Higgs potential in Eq.~(\ref{potential}).  
We can write down the mass matrix of neutrinos in the basis $(\nu_L^{}, N_R^C, S_L^{} )^T$ :
\begin{eqnarray}
{\cal M}=\left( \matrix{0 & Y_\nu^{} v & 0 \cr Y_\nu^T v & 0 & Y_N^{} v_1^{} \cr 0 & Y_N^T v_1^{} & Y_S^{} v_2^{} } \right) \label{inverse}
\end{eqnarray}
where $v, v_1^{}, v_2^{}$ are given in Eq.~(\ref{vev}). 
Given $v_1^{}\sim 1 ~TeV$ and $v_2^{}\sim 1 ~MeV$, the inverse seesaw mechanism is naturally realized.
% we get a natural inverse seesaw model.
The matrix ${\cal M}$ can be diagonalized by the unitary transformation
${\cal U}^\dagger {\cal M} {\cal U}^*= \hat {\cal  M}$; or
explicitly,
\begin{eqnarray}
\left( \matrix{ A & B & C \cr D & E & F \cr G & H & I }
\right)^\dagger \left( \matrix{ 0 & Y_\nu^{} v & 0 \cr Y_\nu^T v & 0 & Y_N^{} v_1
\cr 0 & Y_N^T v_1 & Y_S^{} v_2} \right)  \left( \matrix{ A & B & C \cr D & E &
F \cr G & H & I } \right)^* =\left ( \matrix{\hat M_\nu^{} & 0 & 0
\cr 0 & \hat M_N^{} & 0 \cr 0 & 0 & \hat M_S^{}} \right ) \; ,
\end{eqnarray}
where $\hat M_{\nu, N, S}$ are $3\times 3$ diagonal matrices.
%where $\hat M_{X}^{}= {\rm diag } \{ \hat M_X^1, \hat M_X^{2}, \hat
%M_{X}^3 \} (X= \nu, N, S)$ denote the mass eigenvalues of neutrinos.
The nine mass eigenstates correspond to three observed light neutrinos $\hat \nu$
and six heavy Majorana neutrinos $\hat S$ and $\hat N$, which pair up to form 
three pseudo-Dirac neutrinos. 
%There are nine mass eigenstates: three of them $(\hat
%\nu_i^{})$ correspond to the observed light neutrinos, while the
%other three pairs of heavy Majorana neutrinos $(\hat S_i^{}, \hat
%N_i^{})$ combine to form three pseudo-Dirac neutrinos.

Alternatively, the neutrino mass matrix can be block diagonalized 
and the effective Majorana mass matrix of the active neutrinos can be
approximately written as
\begin{eqnarray}
{\cal M}_\nu^{} =M_D^{} M_R^{-1} \mu M_R^{T-1} M_D^T =v^2 v_1^{-2} v_2^{} Y_\nu^{} Y_N^{-1} Y_S^{}  Y_N^{T-1} Y_\nu^T \; .
\end{eqnarray}
The mass eigenvalues of the three pairs of heavy neutrinos are of the order
$M_R^{}$, 
and the mixing between $SU(2)_L^{}$ singlets and doublets
%and the admixture among $SU(2)_L^{}$ singlets and doublets
is suppressed by $M_D^{}/ M_R^{}$. 
In the basis where the flavor
eignestates of the three charged leptons are identified with their mass
eigenstates, the charged-current interactions between
neutrinos and charged leptons turn out to be
\begin{eqnarray}
-{\cal L_{ \rm CC}^{} } ={ g \over \sqrt{ 2 } }\overline{ \ell_L^{
\alpha } } \gamma_\mu^{} P_L^{} \left( A_{ \alpha i }^{} \hat
\nu_i^{} + B_{ \alpha i }^{} \hat N_i^{} + C_{ \alpha i }^{} \hat
S_i^{} \right) + { \rm h.c. } \; .
\end{eqnarray}
Obviously $A$ describes the charged-current interactions
of light Majorana neutrinos, while $B$ and $C$ are relevant to the
charged currents of heavy neutrinos. 
The neutral current interactions between Majorana neutrinos and 
neutral gauge boson or Higgs can be also written down in a similar way. 
%One can similarly
%write out the neutral current interactions between Majorana
%neutrinos and neutral gauge boson or Higgs.

The explicit expression of  $A$ can be obtained by integrating out heavy 
neutrinos and performing the normalization to the light neutrino wave functions. 
So the effective lepton-mixing matrix can be written as 
\begin{eqnarray}
A_{\alpha i}^{} = \left( \delta_{\alpha \beta}^{} - {1 \over 2 }
\left |M_D^{} M_R^{-1} \mu (M_R^{T})^{-1}\right |^2_{\alpha \beta}
-{1 \over 2} \left |M_D^{} M_R^{-1} \right |^2_{\alpha \beta}
\right) U_{\beta i}^{}\; , \label{mixings}
\end{eqnarray}
where $U$ is the standard PMNS matrix. 
Obviously the effective neutrino mixing matrix is not unitary.
%It's clear that the effective neutrino
%mixing matrix is non-unitary.  
The deviation of $A$ from a unitary matrix is proportional to  $|M_D^{} M_R^{-1}|^2$.  
Constraints on the elements of the leptonic mixing matrix, 
combining data from neutrino oscillation experiments and weak decays was studied in Ref.~\cite{nonunitary} .  
So far neutrino mixing angles have all been measured to a good degree of accuracy, and a preliminary hint for a nontrivial value of $\delta$ has also been obtained from a global analysis of current neutrino oscillation data. But the constraint on the non-unitarity of the lepton mixing matrix still need to be improved and the future neutrino factory can measure this effect through the ``zero-distance" effect and extra CP violations. The Daya Bay~\cite{An:2012eh} reactor neutrino experiment has measured a nonzero value for the neutrino mixing angle $\theta_{13}$ with a significance of 5.2 standard deviations. For this case, even though the neutrino mixing matrix $U$, which diagonalizes the active neutrino mass matrix, takes the well-known lepton mixing pattens, such as  Tri-Bimaximal~\cite{tri} , Bimaximal~\cite{bi} and Democratic~\cite{minzhu} pattens, where $\theta_{13}$ is exactly zero, it is still possible to get relatively large $\theta_{13}$ from the non-unitarity factors in eq.  (\ref{mixings})~\cite{Chao:2011sp}. One can also check the non-unitary effect from the lepton-flavor-violating SM Higgs decays, which,   interesting and important but beyond the scope of this paper, will be shown somewhere else.

%We may find from the expression that the active neutrino mainly mix with the $S_L$, while the mixing with $N_R^C$ is heavily suppressed by the factor $\mu M_R^{-1}$.

%%%%%%%%%%%%%%%%%%%%%%%%%%%
%\begin{figure}[h!]
%\includegraphics[width=8cm]{relic_dm_channel}
%\includegraphics[width=8cm]{relic_dm_h1}
%\includegraphics[width=8cm]{relic_dm_higgs}
%\includegraphics[width=8cm]{relic_dm_lmd3}
%\caption{Relic abundance}\label{fig:relic_dm_channel} 
%\end{figure}
%%%%%%%%%%%%%%%%%%%%%%%%%%%

\section{Dark Matter}

\begin{figure}[t]
\centering
\includegraphics[width=0.25\textwidth]{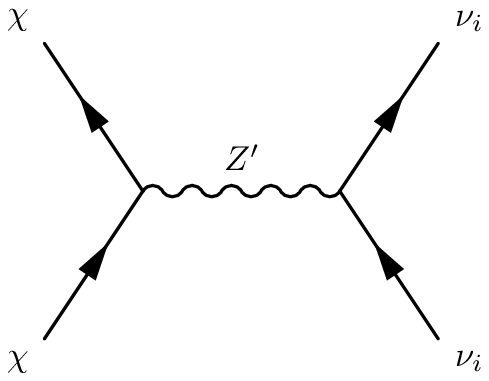}
\hspace{0.5cm}
\includegraphics[width=0.25\textwidth]{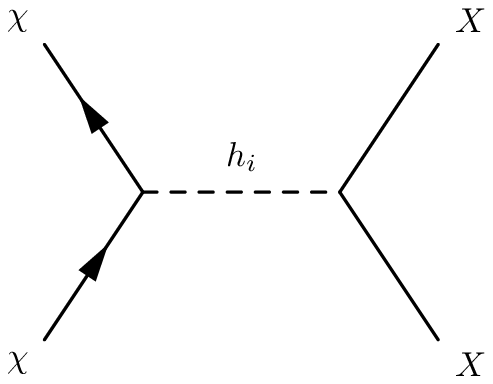}
\hspace{0.5cm}
\includegraphics[width=0.25\textwidth]{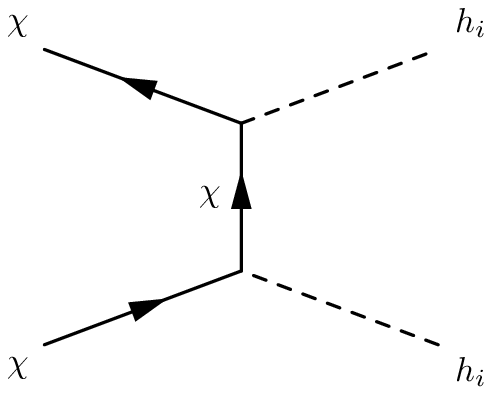}
\caption{Feynman diagrams relevant for the annihilations of the dark matter}\label{fig:anndiag} 
\end{figure}

Precise cosmological observations have confirmed the existence of the non-baryonic cold dark matter.
The lightest generation of $\chi_{L,R}$, the only odd particles under the global $U(1)$ symmetry, can be a
stable dark matter candidate.
%Because of the global $U(1)$ symmetry, $D$, as the only particle 
%with odd parity, can be a stable dark matter candidate. 
In order to produce the dark matter relic abundance observed today
$\Omega_{DM} h^2= 0.1187\pm 0.017$~\cite{Ade:2013zuv}, 
the thermally averaged annihilation rate $\sigma_A v$ should approximately be
 $3\times 10^{-27} cm^3 s^{-1}/\Omega_{DM} h^2$.  
 Interactions relevant to dark matter phenomenology can be written as
\begin{eqnarray}
&&\mbox{I}~~~.  \hspace{2cm}\bar {\chi} \gamma^\mu P_R^{} \chi Z^\prime_\mu    \; , \\
&&\mbox{II}~~.\hspace{2cm}\overline{\nu_L^C}  F^2 \gamma^\mu Z^\prime_\mu \nu_L^C \; ,  \label{eqnG}\\
&& \mbox{III}~.     \hspace{2cm} {Y_\chi /\sqrt{2}}  \bar \chi_L (\cos \theta h_1 -\sin\theta h ) \chi_R^{} \; ,
\label{eqnD}
\end{eqnarray}
where $ \theta$ is the mixing angle between the SM Higgs boson and the Higgs singlet.
It's the 1-2 mixing angle of matrix given in Eq.~(\ref{cpeven}). 
$F$ is either  $D$ or $G$, the $ 21$ and $31$ entry in $\cal {U}$.  The  expressions of $D$ and $G$ can be written as
\begin{eqnarray}
&&G\approx (M_R^{-1 })^* M_D^\dagger U  \; , \\
&&D\approx  (M_R^*)^{-1} \mu^\dagger (M_R^\dagger )^{-1} M_D^\dagger \; ,
\end{eqnarray} 
from which it's easily seen that the active neutrinos mainly mix with $S_L$, while the mixing
with $N_R^C$ is highly suppressed by the factor $\mu M_R^{-1}$. The major contributions to the annihilation cross section come from 
two types of channels,
\begin{equation}
\chi\bar{\chi}\rightarrow Z'\rightarrow 2\nu
\qquad
\chi\bar{\chi}\rightarrow h_i\rightarrow 2X, 
\end{equation}
where $X$ represents the SM fields including $h_i$ but other than neutrinos.
%However, as long as $Z'$ is heavy enough, the contribution to the dark matter annihilation from this channel would be negligible which
%can is clearly shown in the right panel of Fig.~\ref{fig:L3Contour}.
The relevant Feynman diagrams for dark matter annihilation are given in Fig.~\ref{fig:anndiag}. Obviously the dark matter in our model is the hybrid of neutrino portal and Higgs portal. 
%%%%%%%%%%%%%%%%%%%%%%%%%%%
%%%%%%%%%%%%%%%%%%%%%%%%%%%

%%%%%%%%%%%%%%%%%%%%%%%%%%
\begin{figure}[t]
\includegraphics[width=0.45\textwidth]{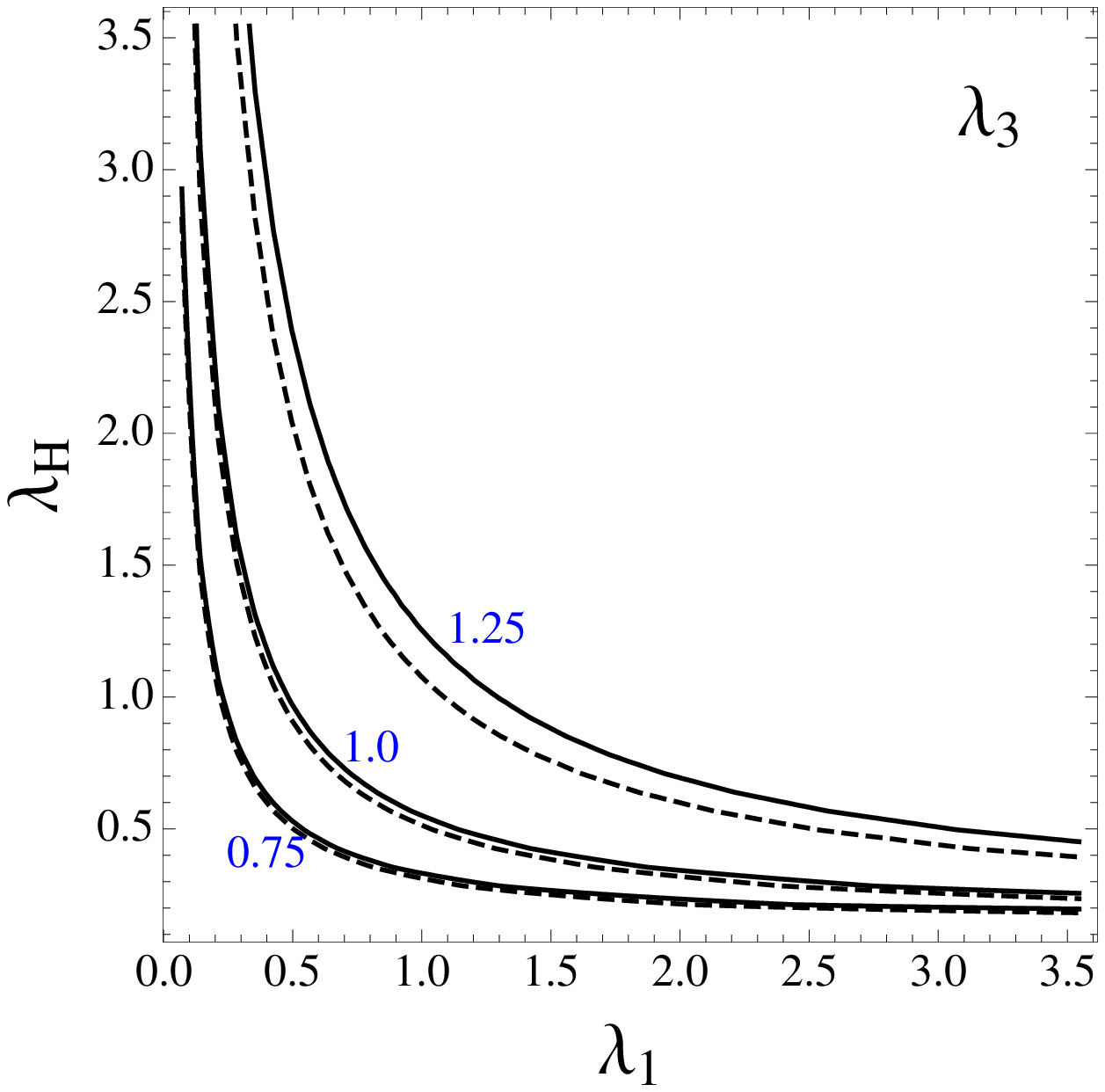}
\hfill
\includegraphics[width=0.45\textwidth]{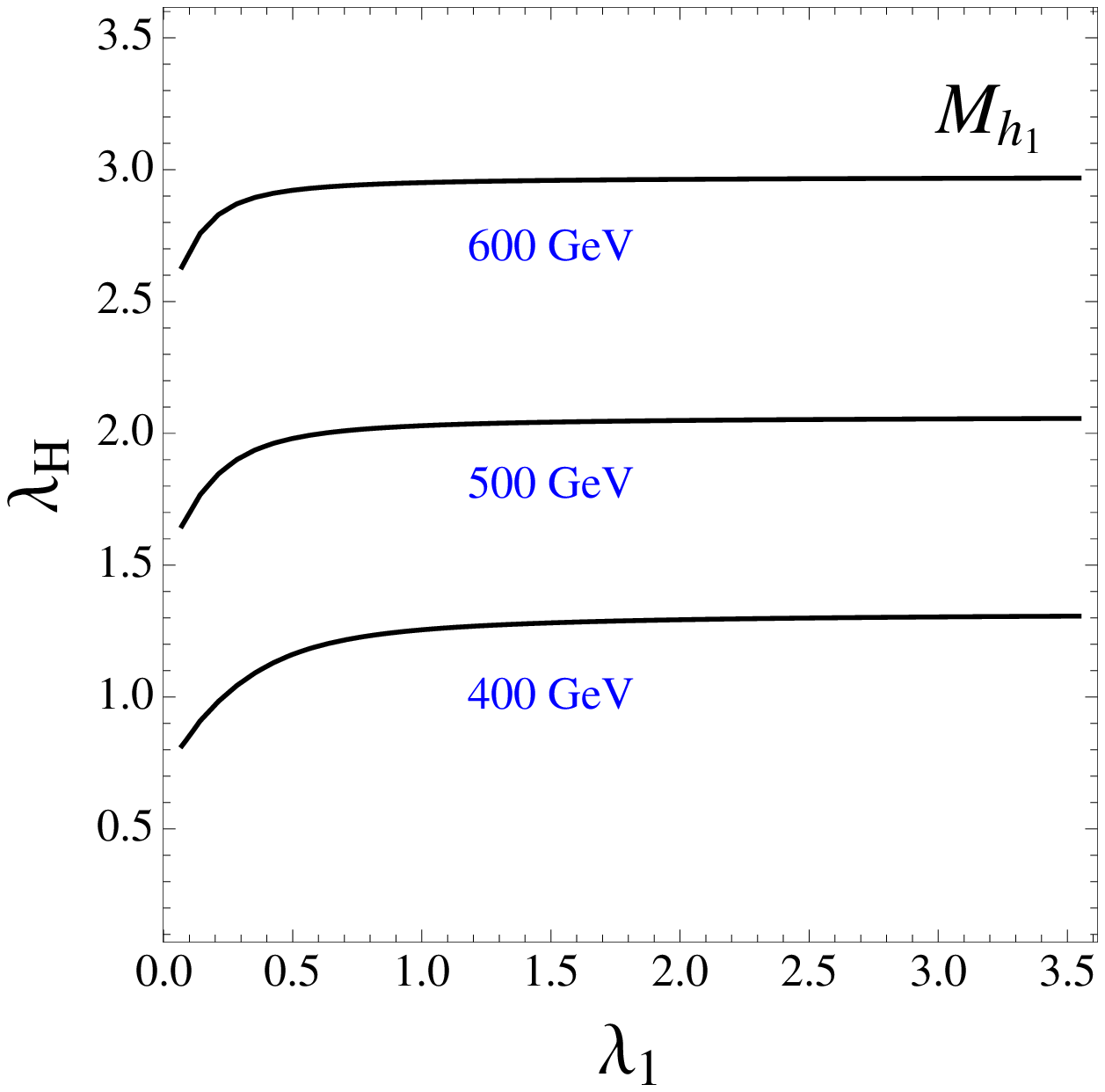}
\caption{Left panel: Contour plot of $\lambda_3$ as a function of $(\lambda_H, \lambda_1)$ which gives a Higgs mass of $126$($120$) GeV
in solid(dashed) lines.
The approximate range for $\lambda_3$ is roughly between $0.5$ and $1.5$ for value chosen in Table.~\ref{tab:values}.
Right panel: Contour plot of the mass of $h_1$ as a function of $(\lambda_H, \lambda_1)$ which gives the right SM-like Higgs mass. }
\label{fig:L3Contour}
\end{figure}
%%%%%%%%%%%%%%%%%%%%%%%%%%%

To investigate the viability of this model of providing a good dark matter candidate,
we fix those parameters irrelevant to the dark matter properties and vary the others. 
Without loss of generality we also simplify the calculation by taking diagonal Yukawa coupling matrices,
which are relevant for the generation of neutrino mixing but irrelevant for the dark matter phenomenology.
The typical input parameters are given in the table. \ref{tab:values}. The relics density and direct detection cross section are calculated with {\it micrOMEGAs}\cite{MicO}, which solves the Boltzmann equations numerically and utilizes {\it CalcHEP}~\cite{calchep} to calculate the relevant cross section.  
\begin{table}[t]
\centering
\begin{tabular}{c|c||c|c}
\hline \hline 
parameters &  values or range &  parameters & values or range  \\
\hline
$m_1^2/{\rm GeV}$ & $5.0\times 10^4$ & $m_2^2/{\rm GeV}$ & $1.0\times 10^6$\\
$Y_\nu,~Y_N$ &0.5  & $Y_S$ & 0.5\\
  $\lambda_H,~\lambda_1$ & $(~0,~\sqrt{4\pi}~]$  & $\lambda_2,~\lambda_4,~\lambda_5$ & 0.5 \\
$M_{Z^\prime}/{\rm GeV}$ & $200,1000$ &$M_\chi/{\rm GeV}$ & [~10,~2000] 
  \\
\hline \hline
\end{tabular}
\caption{ Input parameters at the benchmark point. The parameters in the right part of the table do not change the DM relic density. $\lambda_3$ is calculated by imposing the condition $M_h=126 \ \rm{GeV}$.  The choice of parameter space ensures $v_1$ is of $\rm{TeV}$ and $v_2$ is of $\rm{GeV}$ to generate the right neutrino mass scale. } \label{tab:values}
\end{table}
We show in the left panel of Fig.\ref{fig:L3Contour} the contour plot of $\lambda_3$ as a function of $(\lambda_H, \lambda_1)$ with $m_h=120, 126$ GeV . The approximate range for $\lambda_3$ is roughly $(0,5, 1.5)$ for value chosen in Table.~\ref{tab:values}.
We also show in the right panel of Fig.\ref{fig:L3Contour} the contour plot of the CP-even exotic Higgs mass $M_{h_1}$  as a function of $(\lambda_H, \lambda_1)$ with
the SM-like Higgs mass fixed at $126 \ \rm{GeV}$, which shows that the mass of the exotic CP-even Higgs is in the range of $300-700 \ \rm{GeV}$.

%%%%%%%%%%%%%%%%%%%%%%%%%%%
\begin{figure}[h!]
\centering
\subfloat[]{\label{anbrlight}
\includegraphics[width=0.45\textwidth]{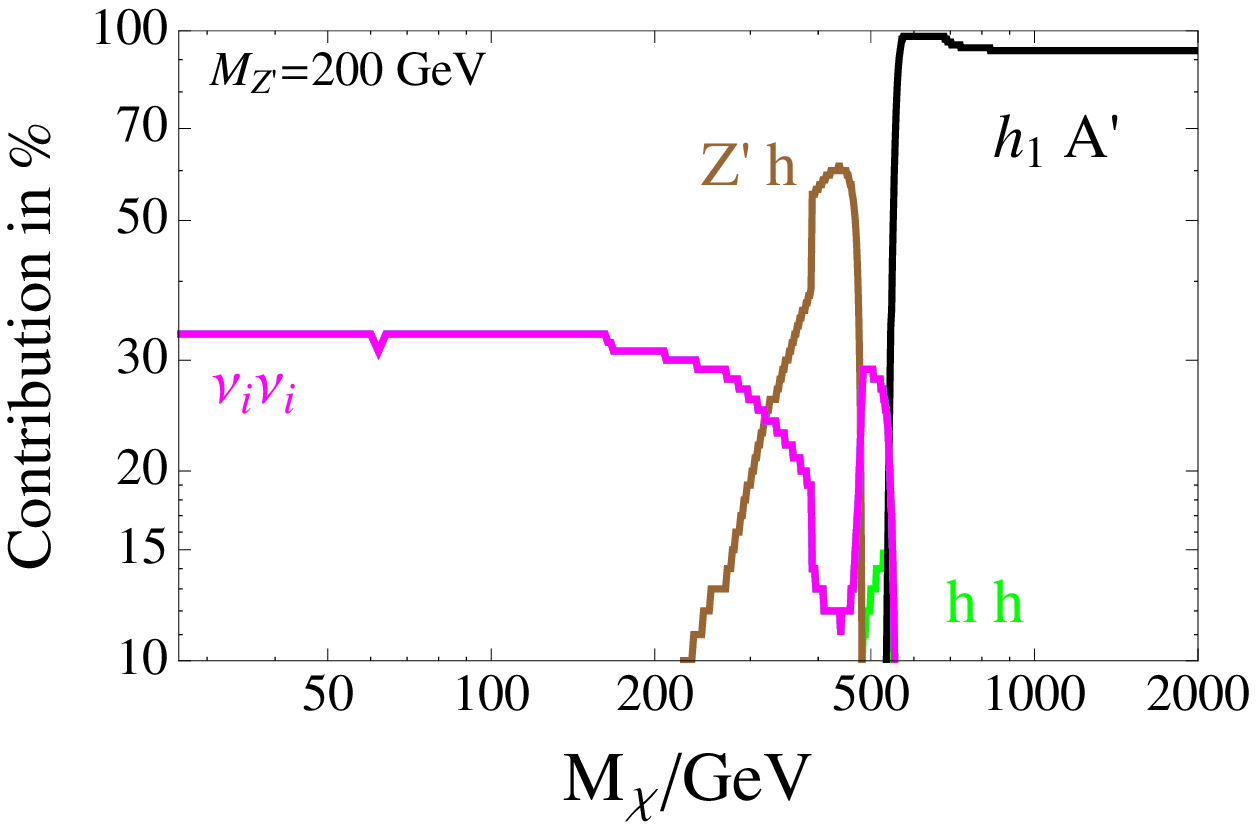}}
~\subfloat[]{\label{anbr}
\includegraphics[width=0.45\textwidth]{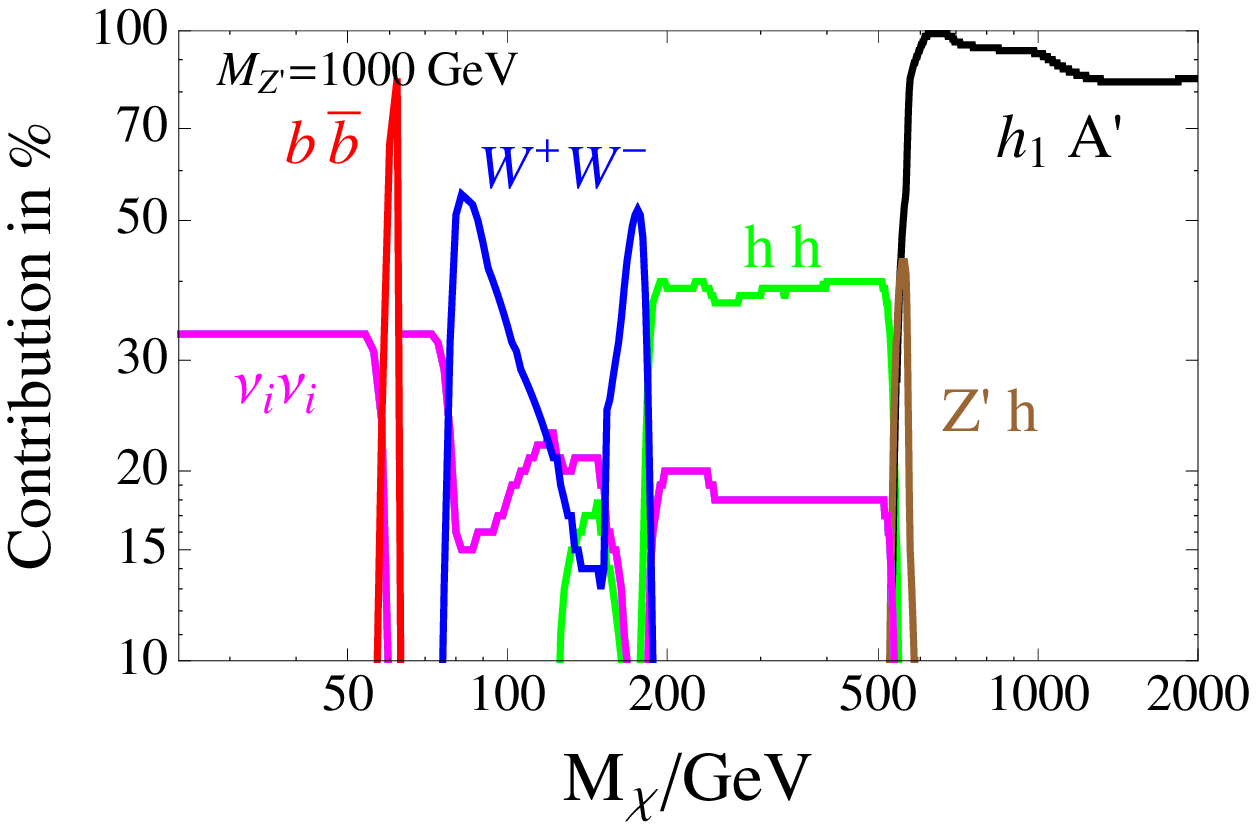}}
\caption{
Relative contributions of different channels to $\Omega_\chi^{-1}$ as a function of dark matter mass
shown in $\%$ for $M_{Z'}=200 \ \rm{GeV}$ (left) and $M_{Z'}=1000 \ \rm{GeV}$ (right)
with $\lambda_H=\lambda_1=1.0$ and other values taken according to Table~\ref{tab:values}.
The channels with less important contributions are not plotted here.  
 }\label{fig:dmrelic}
\end{figure}
%%%%%%%%%%%%%%%%%%%%%%%%%%%

Fig.~\ref{fig:dmrelic} shows the major contributions of various channels to the dark matter annihilation for different parameters.  
Fig. ~\ref{anbrlight} is for $M_Z^\prime =200~{\rm GeV}$ and fig.~\ref{anbr} is for $M_Z^\prime =1~{\rm TeV}$.
For $M_\chi \lesssim M_W$, the dark matter pair annihilate mostly into quark and lepton pairs,
the amplitude of which is suppressed by the Yukawa couplings.
As a result, the dark matter relic abundance for $M_\chi\lesssim M_b$ will be too big to be consistent with the observation. For $M_W \lesssim M_\chi \lesssim M_{h}$, the dominant channels are 
$\chi\bar{\chi}\rightarrow W^+W^-$ and $\chi\bar{\chi}\rightarrow ZZ$. 
When $M_\chi$ gets even bigger, $\chi\bar{\chi}\rightarrow h h$  and $\chi\bar{\chi}\rightarrow h_1 h_1$ are no longer kinematically forbidden and becomes the dominant annihilation channel.  While for $M_\chi > {1/2} (M_{h1}+ M_{Z^\prime})$,  $\chi\bar{\chi}\rightarrow h_1 Z^\prime$ becomes the dominate annihilation channel of the dark matter. For the two examples shown in Fig.~\ref{fig:dmrelic}, the model on the left panel has too big an annihilation cross section which is
excluded and the one on the right will produce the right amount of dark matter given $M_\chi\sim 30 \; \rm{GeV}$.

\begin{figure}[t!]
\centering
\includegraphics[width=0.7\textwidth]{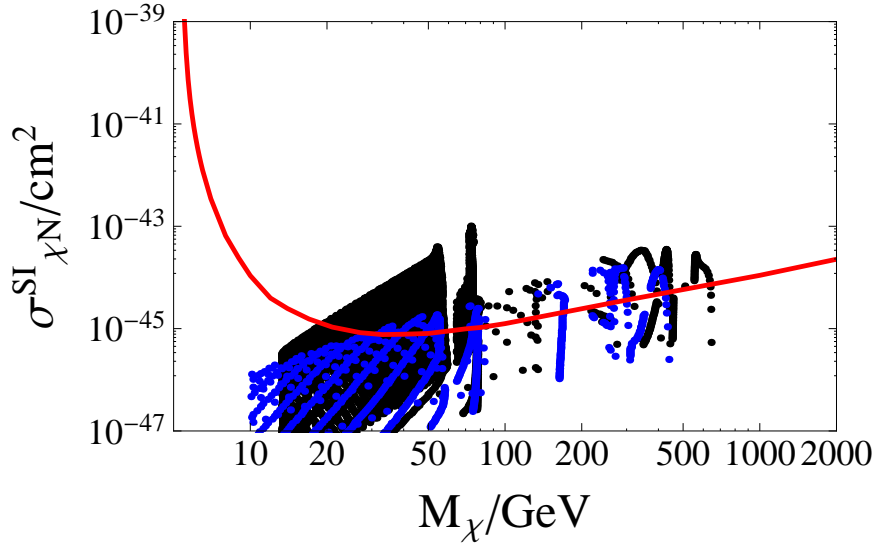}
\caption{Spin-independent nucleon direct detection cross section as a function of the dark matter mass, 
where the  blue and black points are models with $M_{Z'}=200 \ \rm{GeV}$ and $M_{Z'}=1000\ \rm{GeV}$ respectively.
The red solid line is the LUX limit.}
\label{fig:sigma}
\end{figure}

Dark matter is also constrained by direct detection experiments such as LUX ~\cite{Akerib:2013tjd} and XENON 100\cite{xenon100}.  The dark matter -quark interactions in the effective models naturally induce the dark matter-nucleus interactions.  The effective Hamiltonian in our model can be written as
\begin{eqnarray}
H_{\rm eff}= \sum_q c_\theta s_\theta  { m_\chi \over v_1 }( \bar \chi \chi) \left({1 \over M_h^2 }-{1 \over M_{h1}^2 }\right) {m_q \over v } \bar q q \; ,
\end{eqnarray}
where $c_\theta =\cos \theta$ and $s_\theta =\sin \theta$. Parameterizing the nucleonic matrix element as $\langle N \sum_q m_q \bar q q \rangle =f_N m_N$, where $m_N$ is the proton or neutron mass and $f_N$ are the nucleon form factors. We refer to \cite{dmn1,dmn2,dmn3} for explicit values of $f^{p, n}$. The cross section for the DM scattering elastically from a nucleus is given by
\begin{eqnarray}
\sigma^{\rm SI} = {\mu^2 \over \pi } \left[{ c_\theta s_\theta m_\chi \over v v_1 } \left( {1\over M_h^2} -{1\over M_{h1}^2 } \right)\right]^2 [ Z m_p f^p + (A-Z) m_n f^n ]^2 \label{dmnucl}
\end{eqnarray}
where $\mu = m_\chi m_N/(m_\chi + m_N)$ is the reduced mass of the WIMP-nucleon system, with $m_N$ the target nucleus mass. $Z$ and $(A-Z)$ are the numbers of protons and neutrons in the nucleus. Fig.~\ref{fig:sigma} shows the spin-independent nucleon direct detection cross section as a function of the dark matter mass,  where the  blue and black points are models with $M_{Z'}=200 \ \rm{GeV}$ and $M_{Z'}=1000\ \rm{GeV}$ respectively. The red solid line is the LUX limit. One can see from (\ref{dmnucl}) that the scattering cross section is sensitive to $\lambda_3$, which determines the mixing angle, $\theta$, between the SM-like Higgs and the Heavier scalar singlet.  The direct detection cross section gets bigger when $\lambda_3$ increases.

\section{Concluding remarks}
In this paper we extend the SM with a local and a global $U(1)$ symmetry.  
The smallness of  active neutrino Majorana masses is explained  by the modified inverse seesaw mechanism.  Extra fermion singlets introduced to cancel anomalies of the model can play the role of dark matter.  
Constraints on the model parameter space from dark matter relic density as well as dark matter direct searches are studied.   All the fermion masses  arise from the spontaneous breaking of local gauge symmetries, which is a very appealing feature of the model in the era of Higgs physics.
% which is the characteristic of our model.  This feature is very appealing  in the era of Higgs physics, into which we are entering. 

\begin{acknowledgments}

\end{acknowledgments}
Y. C. was supported in part by Australian Research Council. 
W. C. was supported in part by DOE Grant DE-SC0011095

\end{document}